\documentclass[twocolumn,showpacs,preprintnumbers]{revtex4}

\usepackage{graphicx}
\usepackage{bm}

\begin{document}

\title{Magnetic field controlled vacuum charge in graphene quantum dots
with a mass gap}

\author{P. A. Maksym}
\affiliation{Department of Physics and Astronomy, University of Leicester, 
Leicester LE1 7RH, UK}

\author{H. Aoki}
\affiliation{Department of Physics, University of Tokyo, Hongo, Tokyo
 113-0033, Japan}

\date{\today}

\begin{abstract}
The effect of a magnetic field on the charged vacuum is investigated. The
field dependence of the energy levels causes jumps in the total vacuum
charge that occur whenever an energy level crosses the Fermi level and this
leads to re-entrant cycles of vacuum \textit{charging and discharging}. In
atomic systems these effects require astrophysical magnetic fields of
around $10^8$ T but in graphene with a mass gap they occur in laboratory
fields of about 1 T or lower. It is suggested that an electrostatic graphene
quantum dot
defined by a gate electrode provides a solid state model of the as yet
unobserved charged vacuum as well as a model of an atomic system in an
extreme astrophysical environment. Phase diagrams are computed to show how
the total vacuum charge depends on the confining potential strength and
applied magnetic field. In addition the field dependence of the vacuum
charge density is investigated and experimental consequences are discussed.

\end{abstract}


\maketitle

The long-sought-for charged vacuum \cite{PieperGershtein69,Greiner85} is
the ground state
of strong field quantum electrodynamics (QED). Usually the vacuum is
neutral but it charges in the presence of an electric field strong
enough to lower a bound state into the negative energy continuum. For
example, when the charge on the nucleus of a hydrogenic atom increases to
beyond $\sim 172$ the 1s state enters the negative energy continuum and if
this happens at constant Fermi level, the vacuum charges. This is
accompanied by spontaneous emission of two positrons which would enable
the effect to be observed if a strong enough field could be created. In
principle, this is possible because the critical charge can be exceeded in
a collision between two uranium nuclei. But the interaction time 
is too short to allow the transition to a charged vacuum to occur
\cite{Ruffini10} and, despite much effort, vacuum charging has not
yet been observed. However, it may be possible to observe it in a
semiconductor analogue.

The band gap of a semiconductor is analogous to the mass gap of an atomic
system and semiconductor quantum dots are analogous to natural atoms
\cite{qd}. In electrostatic quantum dots, electrons are confined by 
an electrostatic potential that is generated by a gate electrode and
replaces the Coulomb potential of a natural atom. Normally the dot is
engineered so that the confined electron energies are just below the
edge of the conduction band. But in a material with a small band gap
it should be possible to use a stronger potential to lower a state into the
valence band and create a charged vacuum analogue.

Any material that allows a state to be lowered into the valence band with a
modest gate voltage is
suitable. One candidate is monolayer graphene on substrates that induce a
gap, for example BN \cite{bn}, Ru \cite{Enderlein10} and controversially SiC
\cite{sic}. Other candidates include semiconducting carbon nanotubes and
narrow gap semiconductors. Graphene is the only
candidate with a Dirac-like energy dispersion and hence the candidate
that provides the most precise analogue of QED. Indeed it has already been
suggested that a charged vacuum occurs in graphene in the presence of a
Coulomb impurity with enough charge \cite{Pereira08} but it is
difficult to vary the impurity charge experimentally. However
\textit{any attractive potential} of sufficient strength charges the
vacuum. This means a graphene quantum dot is an accurate and practical
analogue of the atomic charged vacuum.
In addition, dots made of materials with
non-relativistic energy dispersion may allow studies of
unusual charged vacua whose properties differ from the atomic one.

But the most interesting feature of the quantum dot analogue is that
the charged vacuum is extremely sensitive to an external magnetic field.
A rough estimate of the field needed to cause significant
effects may be obtained by equating the rest mass-energy to the cyclotron
energy. For atomic electrons this gives about $10^{10}$ T, an ultra strong
magnetic field that only occurs in extreme astrophysical environments
such as magnetized neutron stars \cite{Lai01}. In contrast, for graphene
with  $m_0c^2 \sim 100$ meV and $c \sim 10^6$ ms$^{-1}$ the same estimate
gives about 10 T, well inside the laboratory regime.

Moreover, important effects already occur at lower fields. Depending
on their quantum numbers, energy levels both rise and fall as the
magnetic field increases. If the energy of a charge carrying state rises
above the Fermi level, the vacuum \textit{discharges} while it
\textit{charges} if the energy of a state falls below
the Fermi level. This leads to \textit{re-entrant} cycles of vacuum
charging and
discharging. The relevant energy scale for these processes is the depth of
the state in the continuum. For a hydrogenic atom with $Z=172$ this is
about 15 keV \cite{Greiner85} and the field needed to discharge the
vacuum is about $10^8$ T, still in the astrophysical regime, while
for a graphene quantum dot it is $\lesssim 1$ T. These processes do not
seem to have been investigated before and are studied here in the context
of a graphene quantum dot. But the wave equation for a graphene dot is just
the Dirac equation so \textit{exactly} the same physics should occur in atomic
systems in ultra strong magnetic fields. The graphene dot is not only an
analogue of the charged vacuum but also an analogue of atomic physics in
extreme astrophysical environments.

\begin{figure}
\includegraphics[height=2.9cm, angle=0.0]{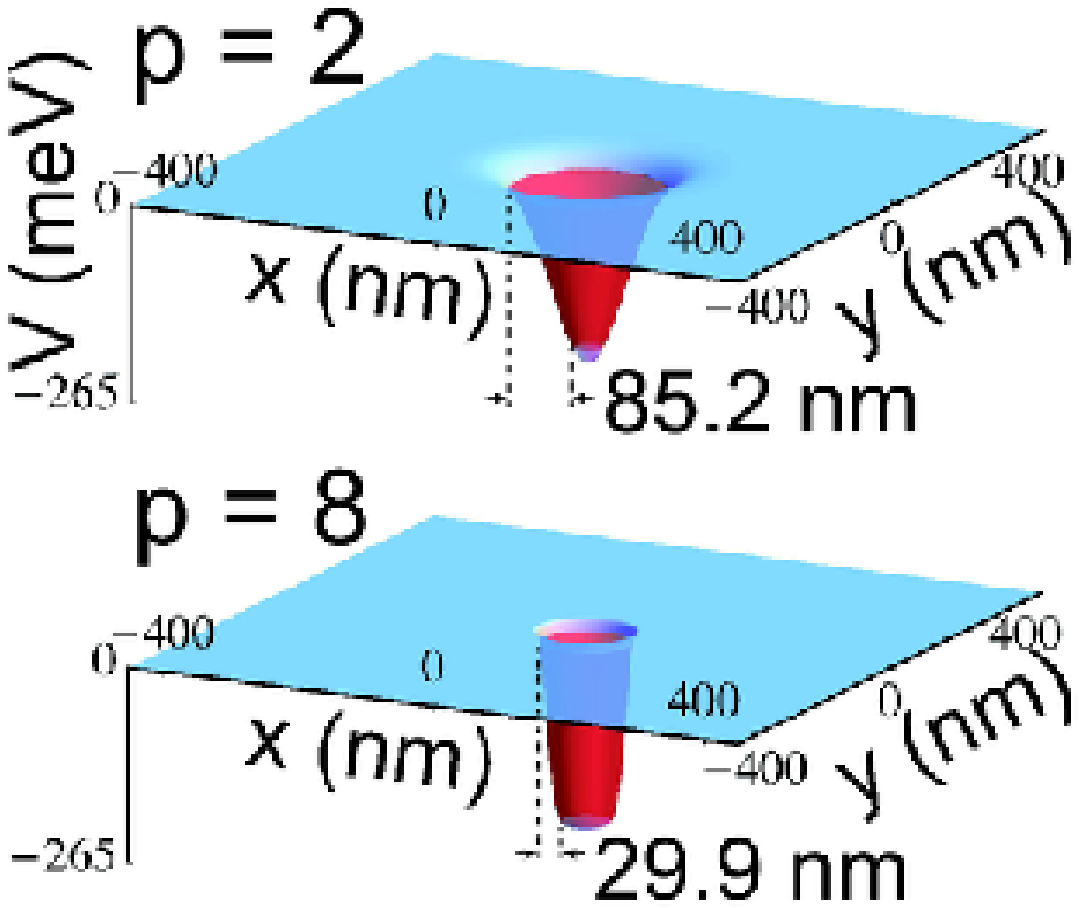}
\includegraphics[height=2.9cm, angle=0.0]{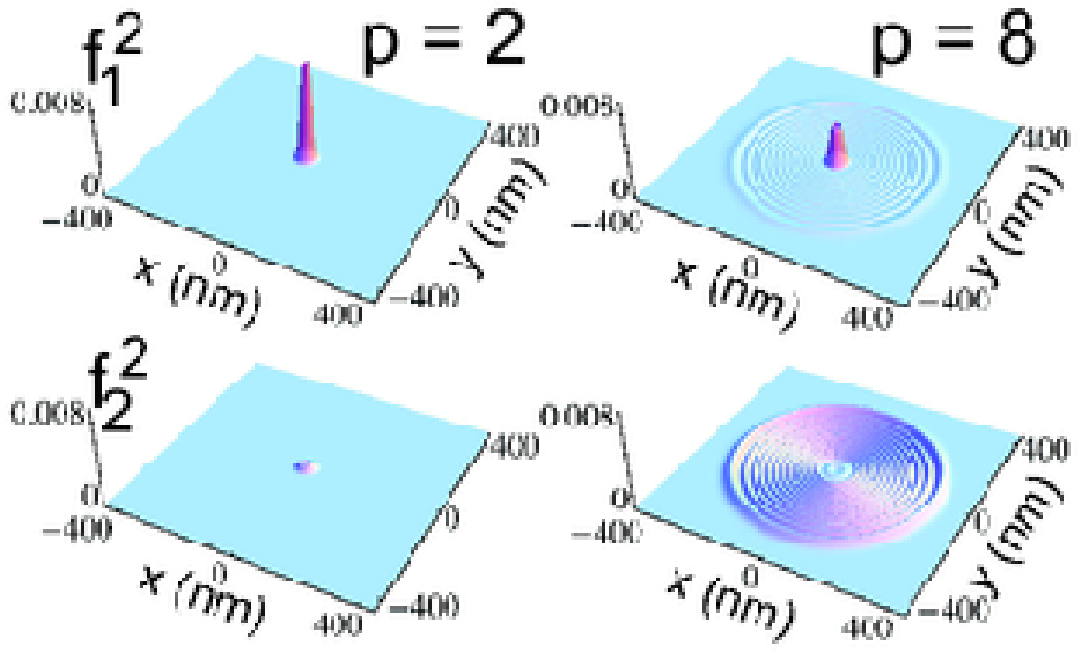}\\
\vspace{5mm}
\includegraphics[height=3.6cm, angle=0.0]{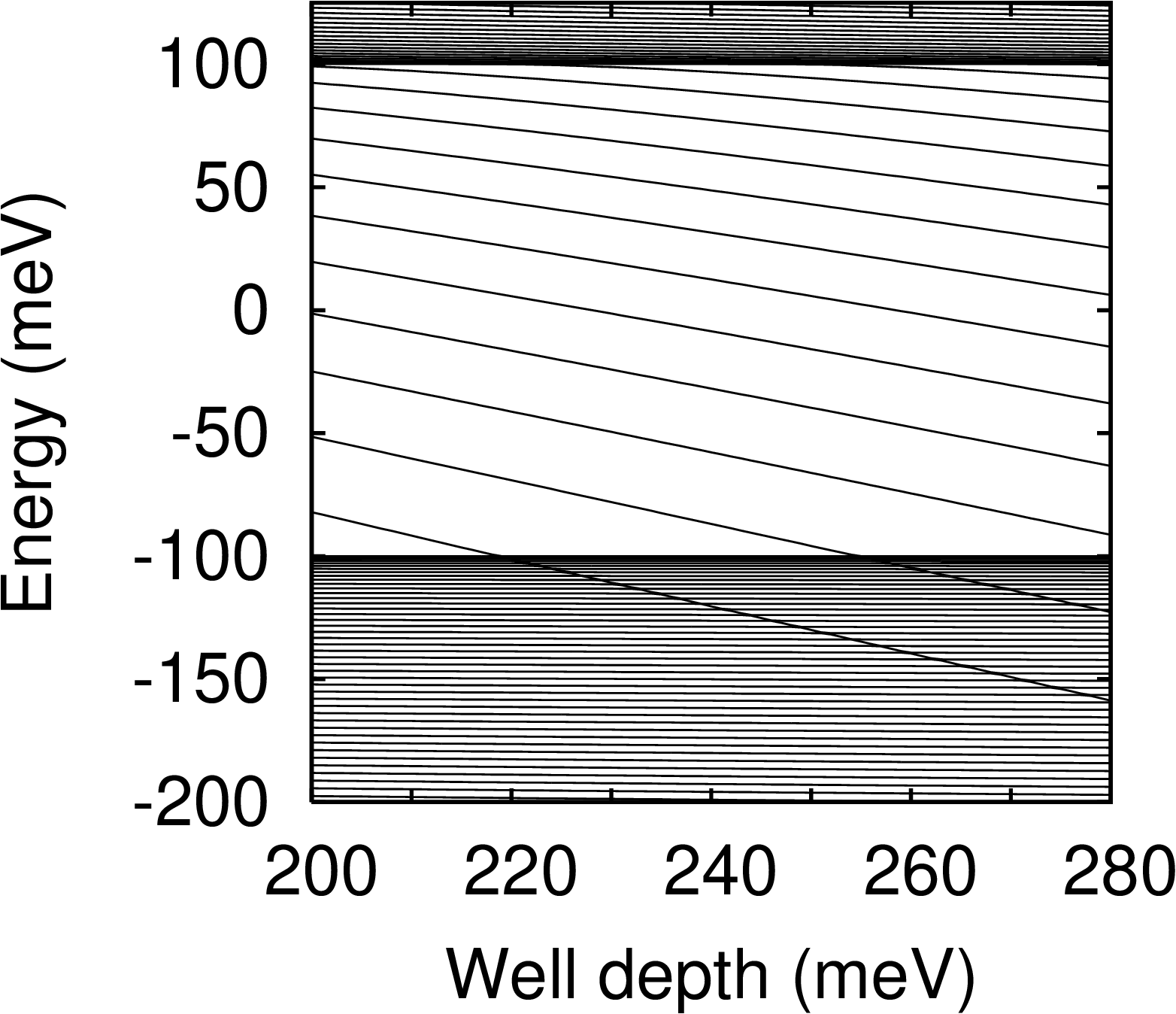}
\includegraphics[height=3.6cm, angle=0.0]{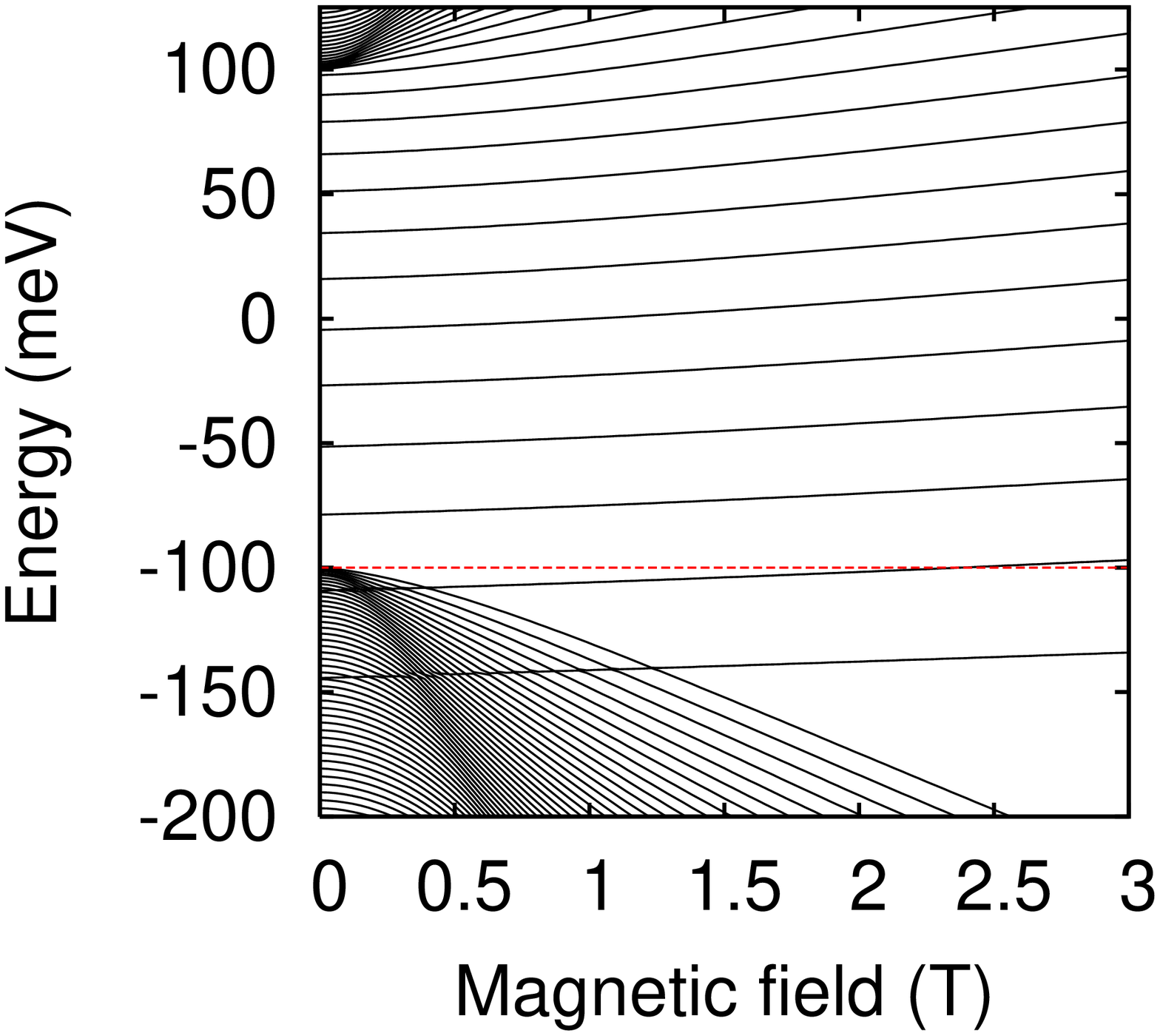}
\caption{Model potentials (top left); well depth and field dependent energy
  levels (bottom)
and typical states at $0.704$ T (top right).}
\label{statefig}
\end{figure}

The objectives of this Letter are, first to demonstrate that charged
vacuum states occur in graphene quantum dots, secondly to demonstrate
magnetic field induced vacuum charging and discharging and finally to
consider the
experimental consequences. The dot is taken to be circularly symmetric
and the electrostatic potential is modelled by
$V(r) = V_0\exp(-(r/\lambda)^p/2)$ where $r$ is the radial co-ordinate,
$|V_0|$ is the well depth and
$\lambda =50$ nm is its width. $p$ determines the shape and slope of the
well; the bottom flattens and the edge sharpens has $p$ increases
(Fig.~\ref{statefig}, top left). The magnetic field, $B$, is taken to be
uniform and perpendicular to the dot plane and
the quantum states are found by solving the two dimensional effective mass
equation. Interestingly, the reduced spatial dimensions may make it easier
to realise
a charged vacuum \cite{Katsura06}. The mass is generated by including a
site-dependent splitting parameter in the Hamiltonian, the same approach as
in earlier work on graphene with a mass gap \cite{Pereira08,Giavaras11}.

The effective mass Hamiltonian for graphene consists of two $2\times 2$
blocks which together are equivalent to the 4-component Dirac Hamiltonian.
One block gives the states near the $K$ point of the Brillouin zone and the
other gives the states near $K'$. The states near $K$ are obtained from the
2-component Hamiltonian $ H =
(\gamma/\hbar) \bm{\sigma} \cdot (\mathbf{p} + e\mathbf{A}) + V
+ m_0c^2 \sigma_z, $ where the $\bm{\sigma}$ are Pauli matrices,
$\mathbf{p}$ is the momentum and $\mathbf{A}$ is the magnetic vector
potential. Here $c = \gamma/\hbar$, $\gamma$ is taken to be 646 meV nm
\cite{Saito98} and $m_0c^2$ is taken to be 100 meV, the upper end
of the observed range (10 - 100 meV).
The eigenstates of a circularly symmetric dot are $\phi(\mathbf{r}) =
(\chi_1(r)\exp(i(m-1)\theta),\chi_2(r)\exp(i m\theta))$, where $\theta$ is
the azimuthal angle
and $m$ is the total angular momentum quantum number.
Equations for the radial functions are obtained by making the substitutions
$f_1 = \sqrt{r}\chi_1$, $if_2 = \sqrt{r}\chi_2$. This leads to
\begin{eqnarray}
\frac{V + m_0c^2}{\gamma}f_1 + \left(\frac{d}{dr} + \frac{m-\frac{1}{2}}{r} +
\frac{e}{\hbar}A_{\theta}\right)f_2 &=& \frac{E}{\gamma}f_1,\nonumber\\
\left(-\frac{d}{dr} +\frac{m-\frac{1}{2}}{r} +
\frac{e}{\hbar}A_{\theta}\right)f_1 + \frac{V - m_0c^2}{\gamma}f_2 &=&
\frac{E}{\gamma}f_2. \nonumber
\end{eqnarray}
These equations are discretized with a second order forward-backward
difference scheme \cite{Giavaras12} which leads to a real,
symmetric eigenvalue problem that is solved numerically. The states near $K'$
are found in a similar way. The numerical system radius is 600 nm.

Physically, the valley index is equivalent to a pseudo-spin and the states
near $K$ and $K'$
correspond to pseudo-spin up and down respectively. The total angular
momentum $\hbar(m-1/2)$ is the sum of the orbital angular momentum and
pseudo-spin. The orbital angular momentum quantum number $l$ is therefore
$m-1$ for states near $K$ and $m$ for states near $K'$.

Figure \ref{statefig} shows the typical behaviour of energy levels and
quantum states. The lower left frame shows the $K$ energy levels as a
function of well depth when $B=0$, $m=1$ (i.e. $l=0$) and $p=2$.
The results clearly show bound state levels
plunging into the negative energy continuum as the well depth increases and
similar behaviour is found for other values $m$ and $p$. The lower right
frame shows the $K$ energy levels as a function of $B$ for the same
parameters as in the lower left frame and well depth $265$ meV. The energy
levels in the negative energy continuum \cite{continuum}
move down as $B$ increases but those of the bound
states that have entered the continuum move up. The bound state level
closest to the continuum edge  crosses $E=-m_0c^2$ (dashed line) when
$B\sim 2.4$ T and this corresponds to vacuum discharge in a system with
Fermi level, $E_F \sim -m_0c^2$. The top right frame shows typical states.
When a bound state enters the continuum it hybridizes with
the continuum states and forms a Fano resonance \cite{Fano61}.
Individual states that contribute to the resonance closest to the
continuum edge are shown in the figure. The resonance width depends on the
strength of the hybridization.  Semi-classical analysis \cite{Maksym12a}
shows that a forbidden region surrounds the dot. As shown by the arrows in
Fig.~\ref{statefig}, the width of this forbidden region decreases when
$p=8$ and this strengthens the hybridization \cite{Supplement}.
When $p=2$, the resonance width is less than the
numerical continuum level separation ($\sim 1$ meV) and the resonance
consists of one state. 
But when $p=8$, it involves about 2-4 states. In the case of
Fig.~\ref{statefig}, $B$ is chosen so that $f_1$ has the roughly the
same amplitude for the 2 main contributing states and the resonance width
is about 3 to 4 meV. The superposition of bound and continuum
character is clearly visible. In contrast, strongly hybridized states do
not normally occur in
Coulomb potentials because the width of the forbidden region is large.

The vacuum charge density is the charge density induced in response to an
external potential, $V$. It may be found from the
commutator QED charge operator, $(-e/2)[\bar{\psi},\gamma^0\psi]$, where
$\gamma^0$ is a Dirac matrix, $\psi$ is the Dirac field operator
and $\bar{\psi}$ is its adjoint. Or it may be found from the normally
ordered QED charge operator, $\hat{N}(-e\bar{\psi}\gamma^0\psi)$.
Alternatively it is given by the standard form
$\rho(V) - \rho(0)$ where $\rho$ is found by summing over states
below $E_F$. The two QED operators are identical \cite{Bjorken65}.
Further, both are equivalent to the standard form and this follows from
their expectation values. The vacuum expectation value of the
commutator operator is
$\hat{\rho}(\mathbf{r})=(-e/2)
(\sum_{E_n < E_F,\alpha} |\phi_{n\alpha}(\mathbf{r})|^2 -
\sum_{E_n > E_F,\alpha} |\phi_{n\alpha}(\mathbf{r})|^2)$, where $\alpha$
represents the component and valley indices. The
standard form is $
-e\sum_{E_n < E_F,\alpha} |\phi_{n\alpha}(\mathbf{r})|^2 - \rho(0) =
(-e/2) (\sum_{E_n < E_F,\alpha} |\phi_{n\alpha}(\mathbf{r})|^2 -
\sum_{E_n > E_F,\alpha} |\phi_{n\alpha}(\mathbf{r})|^2) + 
[(-e/2)(\sum_{E_n < E_F,\alpha} |\phi_{n\alpha}(\mathbf{r})|^2 +
\sum_{E_n > E_F,\alpha} |\phi_{n\alpha}(\mathbf{r})|^2) - \rho(0)])$.
The terms in the square brackets vanish because of completeness and
chiral symmetry \cite{Maksym12b} so the standard form is
equivalent to the QED forms. $\hat{\rho}$ is used to find the charge
density and its integral gives the total vacuum charge $\hat{Q} = (-e/2)
(\sum_{E_n < E_F,\alpha} - \sum_{E_n > E_F,\alpha})$. In QED the sums in
$\hat{\rho}$ and $\hat{Q}$ are divergent and have to be treated with charge
renormalization but they are convergent in the present numerical model
because the energy spectrum is bounded.

\begin{figure}
\includegraphics[height=4.0cm,angle=-90.0]{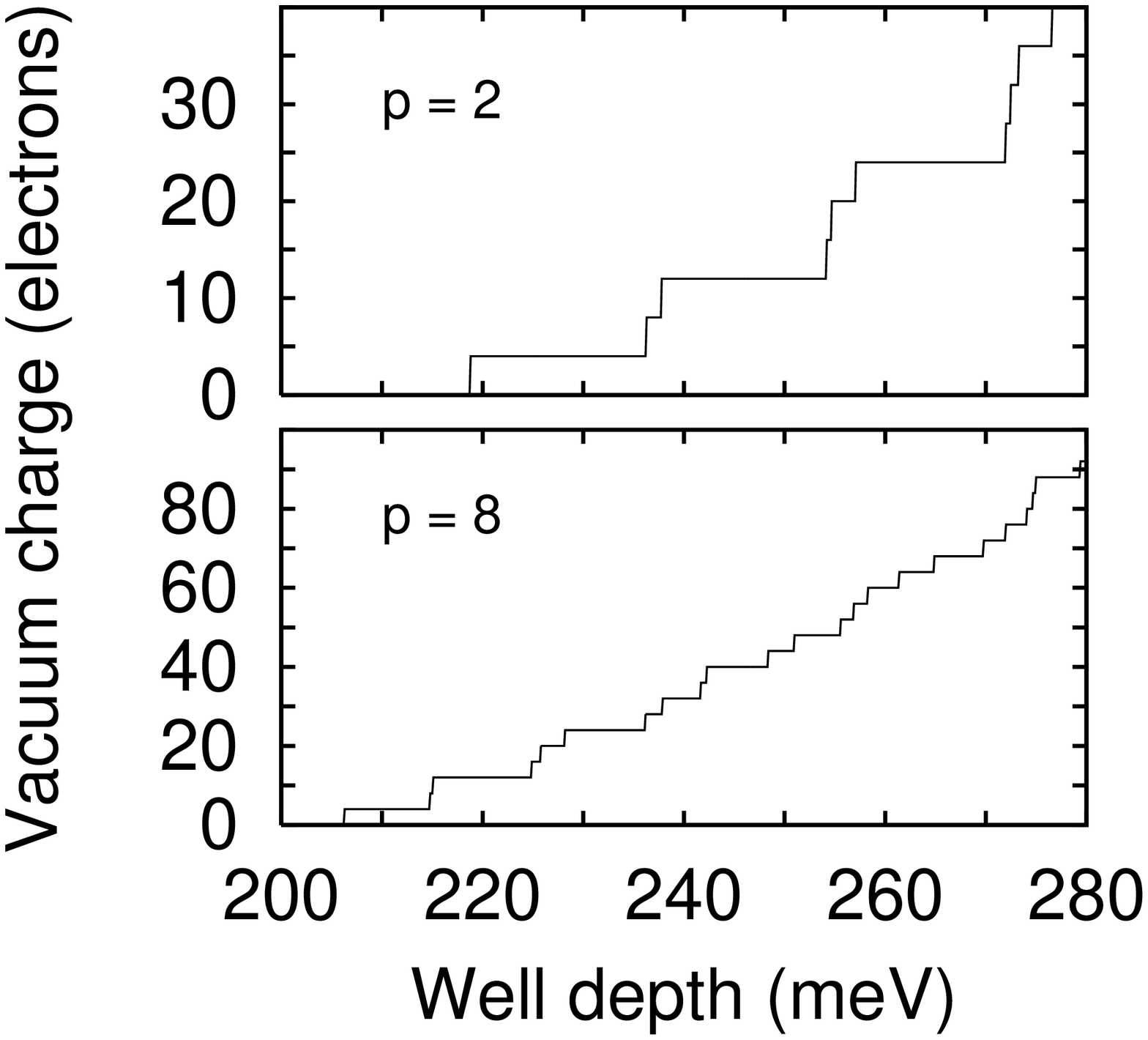}\hspace{3mm}\includegraphics[height=4.0cm,angle=-90.0]{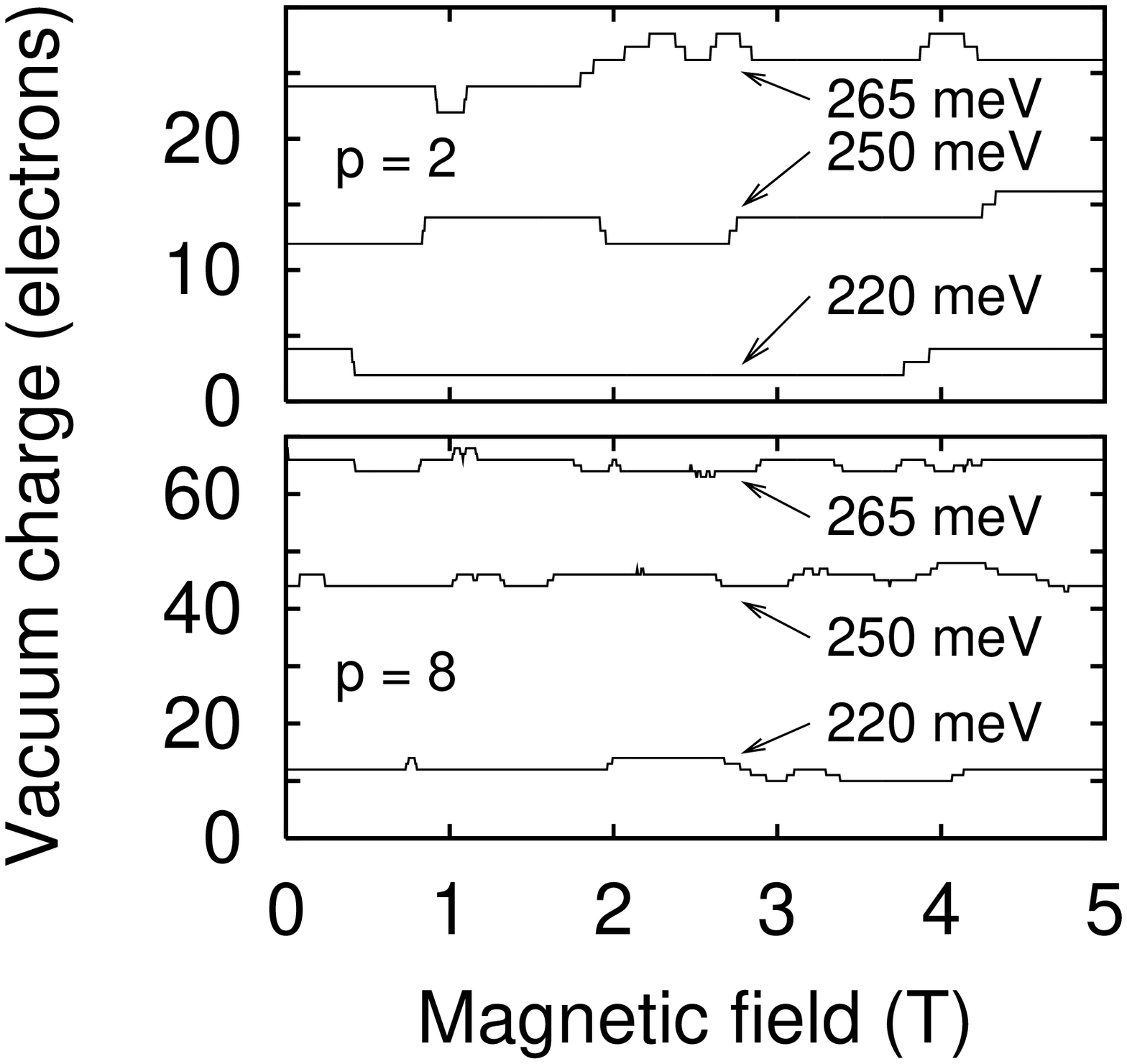}
\caption{Total vacuum charge as a function of well depth (left) and
magnetic field  (right).
}
\label{vcfig}
\end{figure}

The total vacuum charge as a function of well depth and magnetic field
is shown in
Fig. \ref{vcfig}. Real spin splitting is included and the effective
$g$-factor is taken to be 2.0. $E_F$ is just above $-m_0c^2$ so all
real-spin-split Landau levels remain below it.
For $B=0$ the vacuum charge increases monotonically with
well depth in a series of steps. The first step occurs when the $l=0$ level
shown in Fig. \ref{statefig} enters the vacuum. Each step has height 4 and
this corresponds to a 2-fold pseudo-spin degeneracy and a 2-fold real spin
degeneracy. At constant well depth the vacuum charge as a function of $B$
shows re-entrant
behaviour for both values of $p$. For instance, for well depth
$220$ meV and $p = 2$, it falls when $B\sim 0.41$ T and
then rises when $B\sim 3.9$ T. The fall occurs because the $l=0$, $K$
level enters the continuum while the rise 
occurs because the $l=-1$, $K'$ level leaves the continuum. The
behaviour at larger well depth is similar but richer because some levels go
through a minimum as a function of $B$ \cite{Maksym12b}.
Another effect of the field is
real spin splitting. This allows an odd numbered vacuum charge and
leads to double steps of height 1, for example at
$B\sim 3.9$ T with well depth $220$ meV and $p=2$.

\begin{figure}
\includegraphics[height=4.0cm, angle=-90.0]{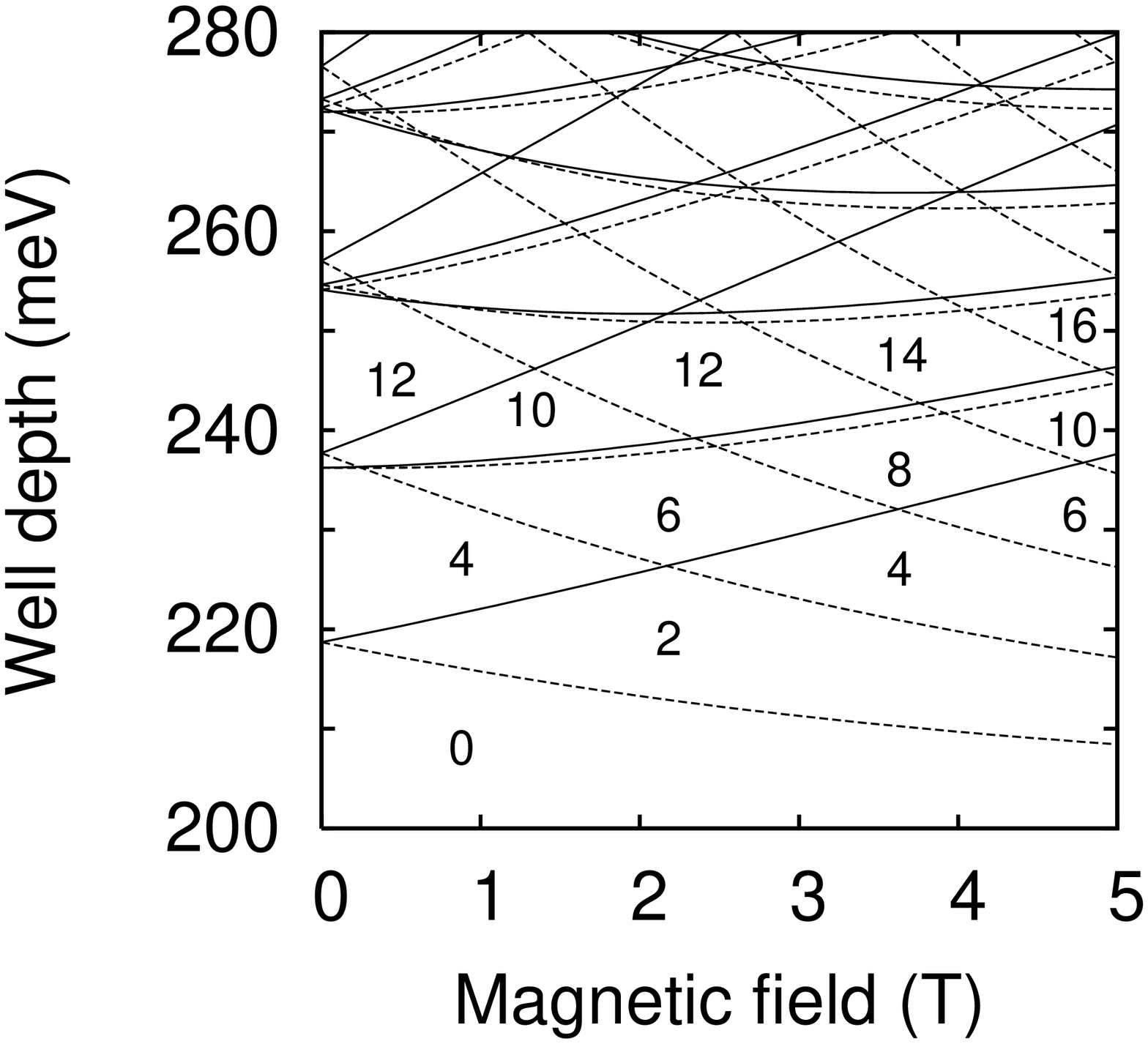}\hspace{3mm}\includegraphics[height=4.0cm, angle=-90.0]{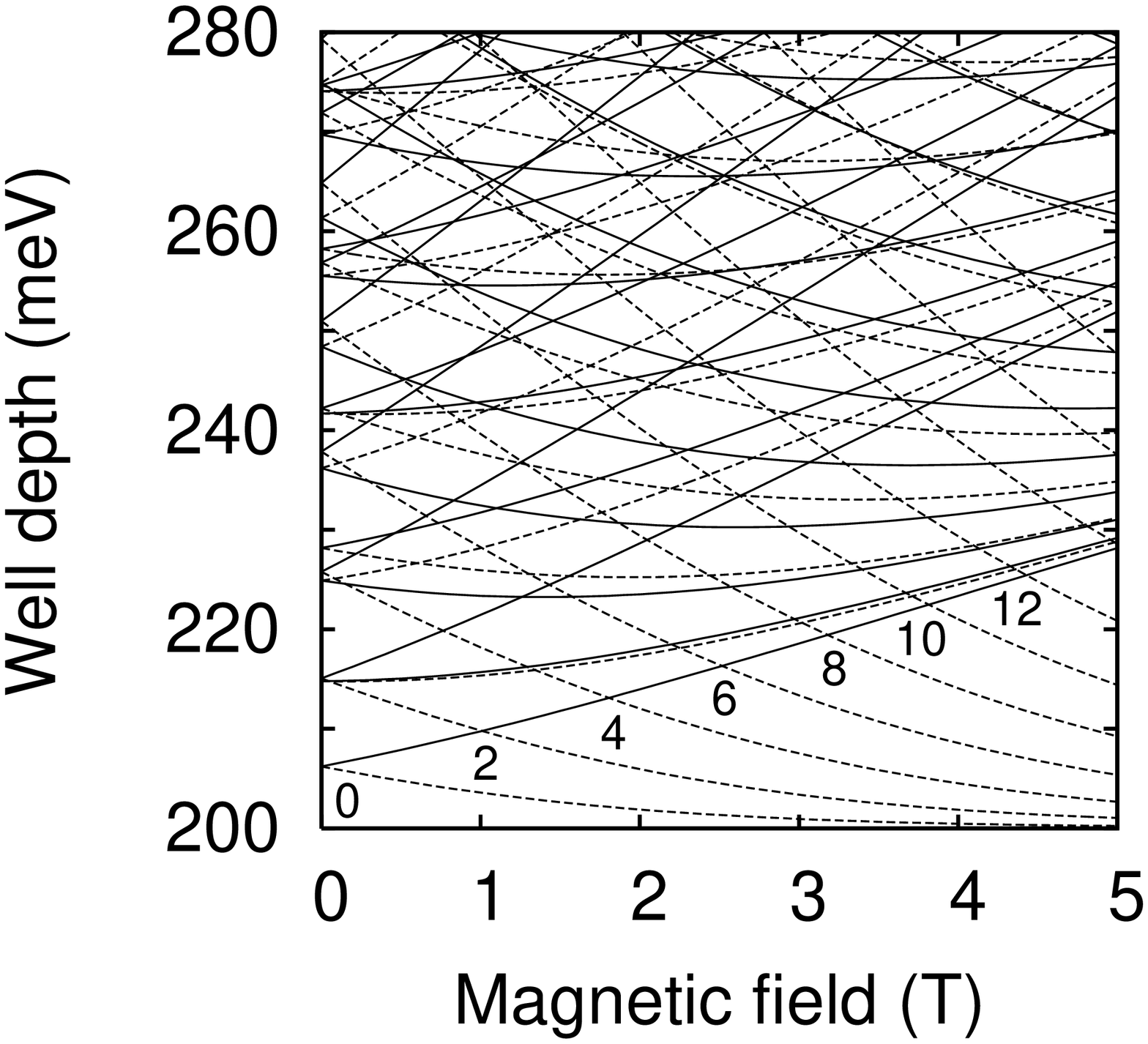}
\caption{Vacuum charge phase diagrams for $p=2$ (left) and $p=8$ (right).
$K$ phase boundaries solid, $K'$ dashed.}
\label{pdfig}
\end{figure}

To investigate the details, vacuum charge
phase diagrams are computed (Fig. \ref{pdfig}). Each phase boundary indicates
where an energy level, $E_n$ crosses the Fermi level and is given implicitly by
$E_n(V_0, B) = E_F$. The total vacuum
charge is shown on selected portions of the diagrams.
For clarity real spin splitting is not
included. Hence each line corresponds to a vacuum charge step of 2
electrons when $B\ne 0$ and 4 when $B=0$. The results in Fig.~\ref{vcfig}
are sections through the phase diagrams with real spin splitting included.
The phase boundaries reflect the physics of the system: they have
have small splittings at $B=0$, unless $l=0$, and there is pronounced
$B$-dependent splitting. These effects can be understood from the
non-relativistic limit
of the effective mass equation. To order $1/m_0^2$ the splitting at $B=0$
results from the pseudo-spin-orbit interaction and the $B$-dependent
splitting results from the interaction of the pseudo-spin with the magnetic
field. Quantitatively, the exact splittings at $B=0$, $V_0=-180$ meV and
$p=2$ are 1.05, 1.88 and 2.49 meV
for the lowest states at $l=1$, 2, 3 respectively, while the lowest order
pseudo-spin-orbit splittings are 1.36, 2.57 and 3.63 meV. The $B$-dependent
splitting at 1 T, calculated from the pseudo-spin $g$-factor is 6.33 meV
while the exact splitting at $l=0$ is 6.05 meV. Hence the non-relativistic
limit describes the physics qualitatively but the exact equation is needed
to find the splittings accurately.

\begin{figure}
\includegraphics[height=4.0cm, angle=-90.0]{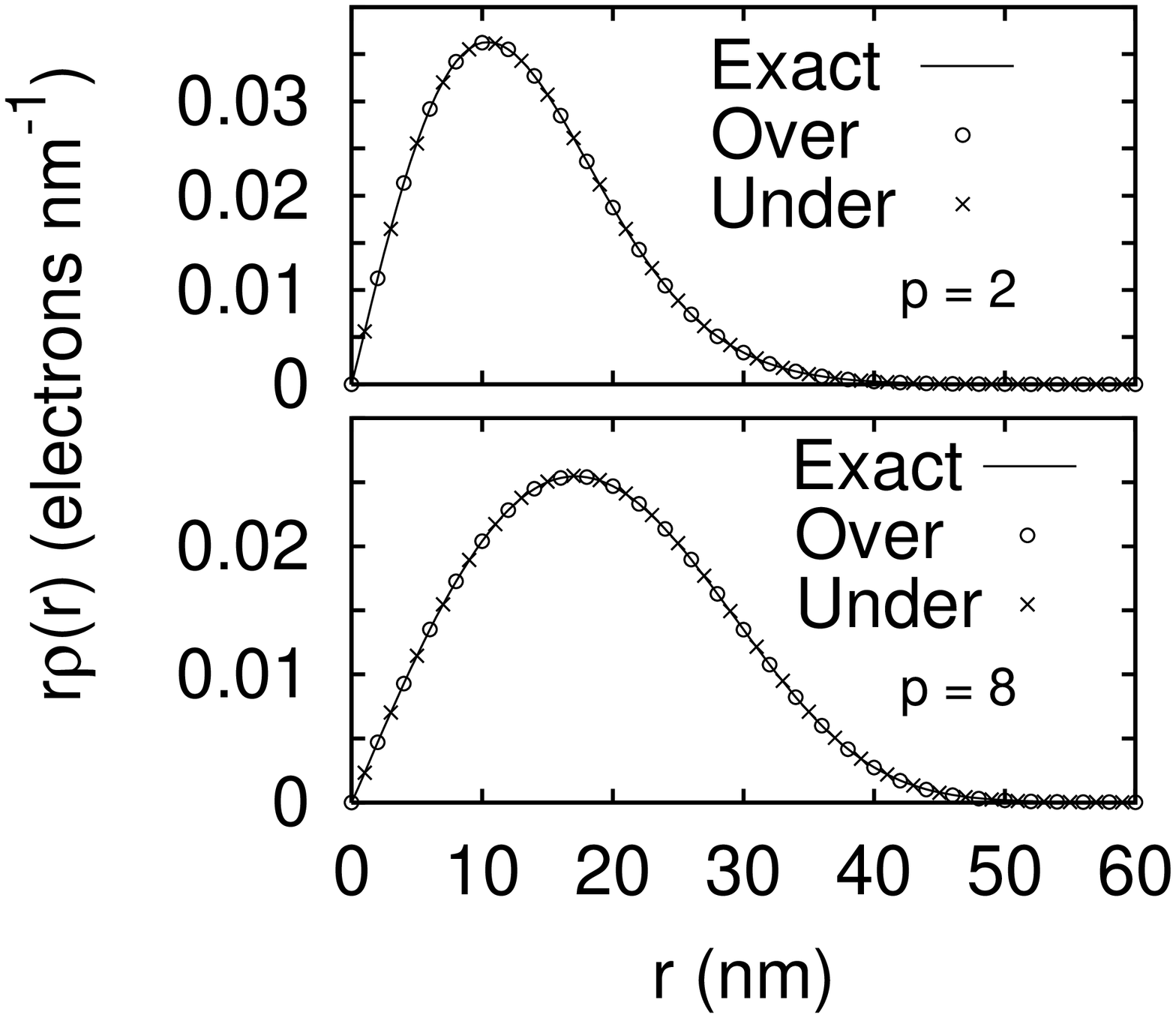}\hspace{3mm}\includegraphics[height=4.0cm, angle=-90.0]{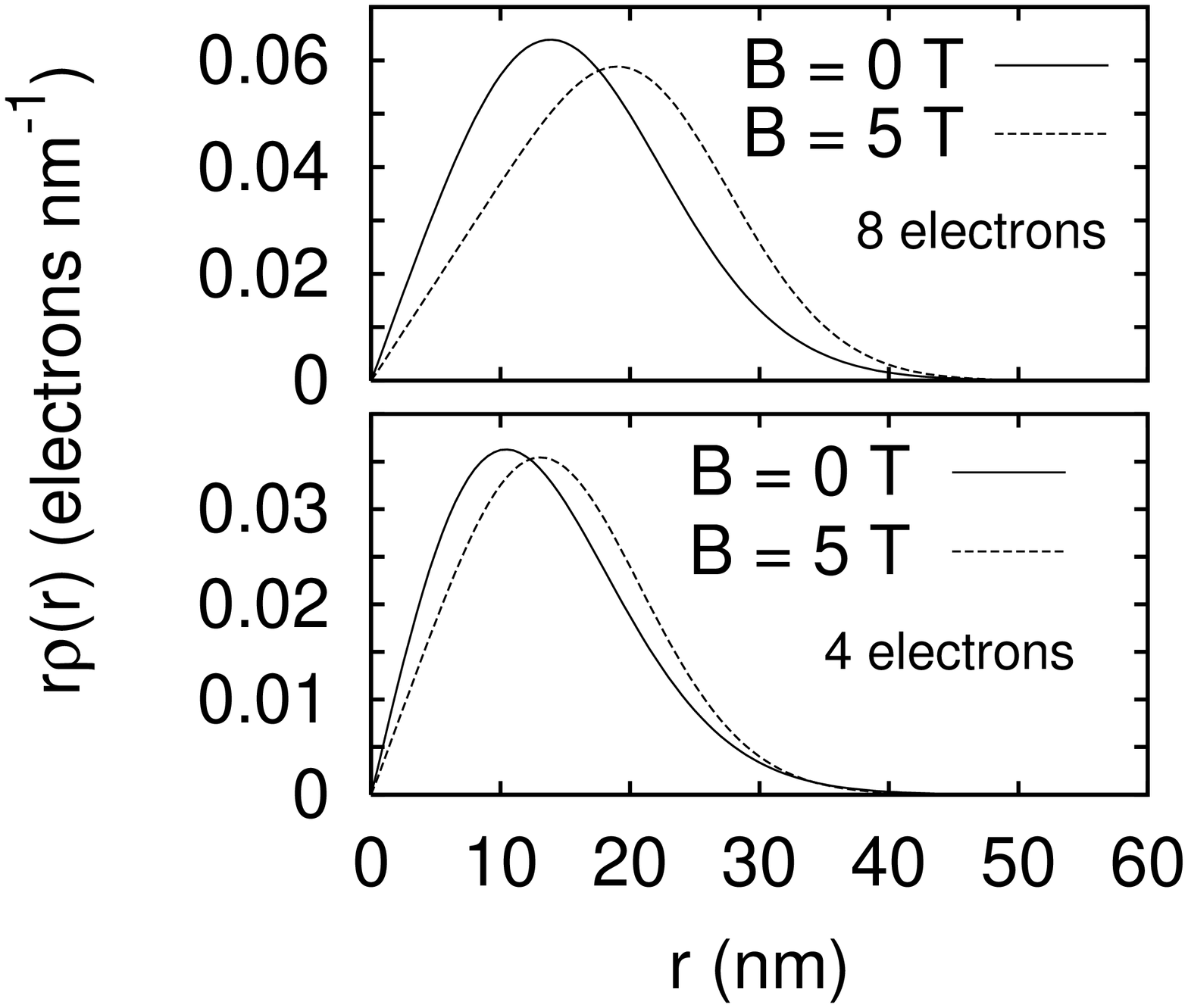}
\caption{Comparison of vacuum charge density and over and undercritical state
  densities (left); magnetic field dependence of $p=2$ vacuum charge
  density for 4 and 8 electrons (right).}
\label{rhofig}
\end{figure}

Vacuum charge densities are shown in Fig.~\ref{rhofig}. The 
charge density increase associated with the charging steps is found from
$\hat{\rho}(V_0+\delta V_0) -\hat{\rho}(V_0)$ where $\delta V_0$ is small
($-0.1$ meV). States below the threshold for vacuum charging are
described as undercritical while those above it are called overcritical
\cite{Greiner85}. The left side of the figure shows
the charge density associated with the first step at $B=0$ T in
Fig.~\ref{vcfig}. To a good approximation, the vacuum charge is stored
in one overcritical state when $p=2$ and two when $p=8$. The open circles
indicate the charge density computed from these few states. Near the peak
they agree with the exact data to better than a few parts in 1000. The
exact density is also well approximated by the undercritical bound state
density (crosses). The $i$\/th overcritical state may be approximated
\cite{Greiner85} by $a_{ib}\psi_b + \sum_c b_{ic} \psi_c$,
where $\psi_b$ is the undercritical bound state and $\psi_c$ are
undercritical negative continuum states. The coefficients $a_{ib}$
and $b_{ic}$ form a unitary transformation \cite{Greiner85} hence the
approximate vacuum charge density reduces to $|\psi_b|^2$. The numerical
data shows this approximation is accurate to a few parts in 1000 near the peak.
A consequence of this approximate sum rule is that the vacuum charge
density is hardly affected by hybridization with continuum states,
although individual states contributing to it are.

The right hand frames of Fig.~\ref{rhofig} illustrate the effect of the
magnetic field on the vacuum charge density when $p=2$ and the total charge
is 4 and 8 electrons. A magnetic field normally compresses the charge
density because the cyclotron length is proportional to $1/\sqrt{B}$.
However, in the present case the density \textit{expands} because states
of higher orbital angular momentum enter the vacuum when the $B$ increases.
These states have a larger spatial extent so the density expands.
For example, with 4 electrons at 0 T the charge density is composed of the
4-fold degenerate $l=0$ state while at 5 T it is composed of real and
pseudo spin split states with $l=0$ and $l=-1$. The expansion continues
until all the higher angular momentum states are exhausted.

There are several experimental consequences of these findings. First,
the vacuum charge in a graphene quantum dot may be detectable via emssion
of holes that is analogous to spontaneous positron emission. It may be
possible to pump the gate voltage to enhance this effect. Secondly, the
strong hybridization may be detectable with scanning tunnelling spectroscopy
(STS). Although the hybridization does not affect the density, the states
in Fig.~\ref{statefig} extend to a radius about an order of magnitude
larger than the dot radius. They should be detectable when the
experimental resolution is less than the width of the
Fano resonance and this requires temperatures of a few K.
Thirdly, it may be possible to detect the vacuum charge directly for
example by capacitance measurements or quantum point contact electrometers.
The ideal material for these experiments should be undoped or p-doped to
ensure that an empty state is taken through the gap and it should be
graphene with a gap to ensure that the effective mass equation is a precise
analogue of the Dirac equation. Existing results \cite{bn,Enderlein10}
suggest that a suitable material can be found. In addition it should be
possible to observe a charged vacuum analogue in narrow gap semiconductors
or semiconducting carbon nanotubes, although the effective mass equation is
no longer relativistic.

In summary, a quantum dot in graphene with a mass gap provides an accurate
model  of the charged vacuum. The vacuum charge is strongly affected by a
magnetic field and a sufficiently high magnetic field can discharge the
vacuum. These effects should also occur for atomic electrons in magnetic
field but the small rest mass-energy of graphene charge carriers,
allows them to be seen with fields $\sim 1$ T instead of the
extreme astrophysical fields needed for atomic electrons. Experimentally, it
may be possible to detect the graphene charged vacuum directly, or by
observation of hole emission or by probing the states with STS. The effect
of interactions remains to be investigated but a Thomas-Fermi approach
\cite{Greiner85} suggests the atomic
vacuum charge survives the effect of interactions.

We thank Seigo Tarucha, Michihisa Yamamoto and Robin Nicholas for useful
discussions. This work was supported by the UK Royal Society and the
Japanese ministry of education, Scientific Research No. 23340112.


\begin{thebibliography}{99}
\bibitem{continuum} In the magnetic field range considered the levels are
  very close to each other and are referred to as a continuum.
\end{thebibliography}
\end{document}